\documentstyle[eqsecnum,preprint,aps,prd]{revtex}
\preprint{WISC-MILW-94-TH-13}
\tighten
\begin{document}
\draft
%%%%%%%%%%%%%%%%%%% This line has 80 characters in it.  %%%%%%%%%%%%%%%%%%%%%%%

%%%%%% defines must be replaced before submitting electronically %%%%%%%%%%%%
\def\i{{\bf \hat i}}
\def\j{{\bf \hat j}}
\def\k{{\bf \hat k}}
\def\vb{{\bar v}}
\def\xb{{\bar x}}
\def\ut{{\tilde u}}
\def\vt{{\tilde v}}
\def\Ut{{\tilde U}}
\def\Vt{{\tilde V}}
\def\zt{{\tilde z}}
\def\oo{ {\omega \Omega}}
\def\idu{ {\int_0^1 du\>}}
\def\idv{ {\int_0^1 dv\>}}
\def\idut{ {\int_0^1 d\ut\>}}
\def\idvt{ {\int_0^1 d\vt\>}}
\def\hti{{\tilde h}}
\def\xt{{\tilde x}}
\def\b{{\bf b}}
\def\a{{\bf a}}
\def\ups{{\upsilon}}
\def\upst{{\tilde\upsilon}}
\def\ap{{\bf a}'}
\def\bp{{\bf b}'}
\def\half{{1 \over 2}}
\def\sumprime_#1{\setbox0=\hbox{$\scriptstyle{#1}$}
\setbox2=\hbox{$\displaystyle{\sum}$}
\setbox4=\hbox{${}'\mathsurround=0pt$}
\dimen0=.5\wd0 \advance\dimen0 by-.5\wd2
\ifdim\dimen0>0pt
\ifdim\dimen0>\wd4 \kern\wd4 \else\kern\dimen0\fi\fi
\mathop{{\sum}'}_{\kern-\wd4 #1}}

%%%%%%%%%%%%%%%%%%%%%%%%%%%%%%%%%%%%%%%%%%%%%%%%%%%%%%%%%%%%%%%%%%%%%%%%%%%%%%%

\title{ANALYTIC RESULTS FOR THE GRAVITATIONAL RADIATION FROM A
CLASS OF COSMIC STRING LOOPS}

\author{Bruce Allen}
\address{
Department of Physics, University of Wisconsin -- Milwaukee\\
P.O. Box 413, Milwaukee, Wisconsin 53201, U.S.A.\\
email: ballen@dirac.phys.uwm.edu}

\author{Paul Casper}
\address{
Department of Physics, University of Wisconsin -- Milwaukee\\
P.O. Box 413, Milwaukee, Wisconsin 53201, U.S.A.\\
email: pcasper@dirac.phys.uwm.edu}

\author{Adrian Ottewill}
\address{
Department of Mathematical Physics\\
University College Dublin, Belfield, Dublin 4, IRELAND\\
email: ottewill@relativity.ucd.ie}

\maketitle
\begin{abstract}
Cosmic string loops are defined by a pair of periodic functions $\a$
and $\b$, which trace out unit-length closed curves in
three-dimensional space. We consider a particular class of loops, for
which $\a$ lies along a line and $\b$  lies in the plane orthogonal
to that line.  For this class of cosmic string loops one may give a
simple analytic expression for the power $\gamma$ radiated in
gravitational waves.  We evaluate $\gamma$ exactly in closed form for
several special cases:  (1) $\b$ a circle traversed $M$ times;
(2) $\b$ a regular polygon with $N$ sides and interior vertex angle
$\pi-2\pi M/N$;  (3) $\b$ an isosceles triangle with semi-angle
$\theta$.  We prove that case (1) with $M=1$ is the absolute minimum
of $\gamma$ within our special class of loops, and identify all the
stationary points of $\gamma$ in this class.

\end{abstract}

%%%%%%%%%%%%%%%%%%%%%%%%%%%%%%%%%%%%%%%%%%%%%%%%%%%%%%%%%%%%%%%%%%%%%%%%%%%%%%%

\section{INTRODUCTION}
\label{section1}

Cosmic strings are topological defects that may have formed at phase
transitions as the universe expanded and cooled \cite{Kibble,%
Zel'dovich,Vilenkin,SV}.  Loops of cosmic string oscillate
and emit gravitational radiation.  This process is of particular
importance because most of the observational limits on cosmic string
networks in the early universe are obtained by considering the effects
of this gravitational radiation (\cite{SV,AllenCaldwell} and references
therein).

The power emitted in gravitational radiation by a cosmic string loop
depends upon its shape and velocity.  In the center-of-mass frame, a
cosmic string loop is specified by the position ${\bf x}(t,\sigma)$
of the string as a function of two variables: time $t$ and a space-like
parameter $\sigma$ that runs from $0$ to $L$.  (The total energy of the
loop is $\mu L$ where $\mu$ is the mass per-unit-length of the string).
When the gravitational back-reaction is neglected, the string loop
satisfies equations of motion whose most general solution in the
center-of-mass frame is
\begin{equation}
{\bf x}(t,\sigma)= {1 \over 2} \big[ \a(t+\sigma) + \b(t
- \sigma)\big].
\label{X}
\end{equation}
Here $\a(u) \equiv \a(u+L)$ and $\b(v) \equiv \b(v+L)$ are a pair of
periodic functions, satisfying the ``gauge condition"
$|\a'(u)| = |\b'(v)|=1$, where $'$ denotes differentiation w.r.t. the
function's argument.  The average power radiated by an oscillating string
loop is given by
\begin{equation}
P = \gamma G \mu^2 c,
\end{equation}
where $G$ is Newton's constant and $c$ is the speed of light.
Quite suprisingly, the dimensionless quantity $\gamma$ depends only
upon the shape of the cosmic string loop determined by the functions
$\a$ and $\b$; it is independent of the length $L$ of the loop
\cite{Vilenkin,AllenShellard}.  From here on, we therefore set $L=1$.

In a recent paper, we presented a new formula for $\gamma$.  The formula
is an exact analytic closed form for any piecewise-linear cosmic string
loop \cite{AllenCasper}.  Although numerical values of $\gamma$ are
easily obtained by that method, the large number of terms makes it
difficult to write out the analytic formula explicitly for most loops.
The conventions and notation used in the present work are adopted from
this earlier work.  In particular, for the rest of this paper we use
units with $c=1$, and use $\i$, $\j$ and $\k$ to denote
a right-handed triad of orthogonal unit vectors in flat $I\!\!R^3$.

In this short paper we evaluate $\gamma$ for a particular class of
cosmic string loops.  For loops in this class, the $\a$-loop traces
out the following path.  When $u$ vanishes, $\a(u)$
also vanishes.  As $u$ increases, $\a(u)$ moves smoothly up the
$z$-axis with unit velocity $\k$, until $u=1/2$.  From $u=1/2$ until
$u=1$, $\a(u)$ moves back down to the origin with unit velocity $-\k$.
For loops in the class that we consider, the $\b$-loop lies entirely
in the $x-y$ plane, i.e. in the plane perpendicular to the path of
the $\a$-loop.  See Figure \ref{fig1} for an example.  Note that the
$\b$-loops in this class are {\it not} required to be piecewise-linear.

The position (\ref{X}) of the cosmic string loop in $I\!\!R^3$ may be easily
visualized.  Consider a string loop at time $t=0$.
  The $x$ and $y$ components of ${\bf x}(\sigma,0)$ are given
by (half) the $x$ and $y$ components of $\b(-\sigma)$.  The $z$ component
of ${\bf x}(\sigma,0)$ is given by (half) the $z$ component of
$\a(\sigma)$.  Thus, when viewed from large positive $z$, the
$x-y$ projection of the string loop looks like the $\b$-loop
(scaled by a factor of $1/2$).  Of course, the string loop does not lie in
a plane.  As $\sigma$ increases from 0 to $1/2$, the $z$ component of
${\bf x}(\sigma,0)$ increases smoothly from 0 to $1/4$.  As $\sigma$
continues to increase from $1/2$ to 1, the $z$ component of
${\bf x}(\sigma,0)$ decreases smoothly from $1/4$
back to 0.  As the value of $t$ changes, the $z$ component of a point on
the string loop oscillates between 0 and $1/4$ while the $x-y$ projection
of the string loop remains unchanged.  From this, it can be seen immediately
that string loops in this class will self-intersect if and only if the
$\b$-loop intersects itself.  In this paper,
self-intersecting string loops are assumed {\it not} to intercommute,
i.e. the strings are assumed to ``pass through" one another without
interaction.

A short outline of the paper is as follows.  In section 2 of the
paper, we derive a simple analytic formula for $\gamma$ for loops in
the particular class described above.  This formula expresses $\gamma$
as a convolution of the $\b$ function with itself.  In section 3 this
formula is exploited to determine $\gamma$ for a several particular
loops.  In the first case, the $\b$-loop is a circle traversed $M$
times in the $x-y$ plane.  In the second case, the $\b$-loop traces out
a regular, $N$ sided polygon with interior vertex angle $\pi-2\pi M/N$
in the $x-y$ plane.  In the third case, the $\b$-loop is an isosceles
triangle in the $x-y$ plane, with semi-angle $\theta$.  In section 4 we
find all of the extrema of $\gamma$ for loops in the limited class
described above.  We prove that the minimum value of $\gamma$ is
attained only by the string loop of the first case with $M=1$, for
which
\begin{equation}
\gamma= 16 \int_0^{2 \pi} {1-\cos x \over x}dx \approx 39.002454.
\label{min}
\end{equation}
This is followed by a short conclusion, in which we also discuss the rate
at which the values of $\gamma$ for the piecewise-linear polygon case with
$M=1$ approach the limiting value (\ref{min}) of $\gamma$ for the perfectly
circular case.

%%%%%%%%%%%%%%%%%%%%%%%%%%%%%%%%%%%%%%%%%%%%%%%%%%%%%%%%%%%%%%%%%%%%%%%%%%%%%%%

\section{FORMULA FOR THE 2-PLANE CASE}
\label{section2}

In this section we find a simple general formula for $\gamma$, valid for
our particular class of cosmic string loops.  The string loops in this
class are defined by $\a$- and $\b$-loops with the following properties.
The $\a$-loop is composed of two straight segments joined at
``kinks" where the tangent vector $\ap(u)$ is discontinuous.
This loop is taken to lie along the $z$-axis with one kink
positioned at the origin.  (The parameter $u$ is chosen to be zero at
this kink.)  The other kink (at $u=1/2$) is positioned above the first kink
and has coordinates (0,0,1/2).  The $\b$-loop is constrained to lie in the
plane perpendicular to the $\a$-loop (i.e., the $x-y$ plane).
The formula for $\gamma$ found in this section will be valid for all cosmic
string loops in this limited class.

The starting point for the calculation in this section is the following
equation,
which is derived in section 3 of reference \cite{AllenCasper},
\begin{equation}
\gamma= 4\idu \idv \idut \int_{-\infty}^{\infty} d\vt\>
\psi(u,v,\ut,\vt) D \left[ \epsilon( \Delta t)
\delta( (\Delta t)^2-|\Delta {\bf x}|^2)\right].
\label{gamma1}
\end{equation}
In this equation, $\delta$ is the Dirac delta function and
$\epsilon(x)=2\theta(x)-1$
where $\theta(x)$ is the unit step function.
The $\ut$ and $\vt$ integrations are over the entire world-tube
swept out in space-time by the string loop as it oscillates.  The
$u$ and $v$ integrations are over a region on the world-tube swept
out by the string loop during a single oscillation.
The functions $\Delta t\equiv (u+v-\ut-\vt)/2$ and
$\Delta {\bf x}\equiv (\a(u)+\b(v)-\a(\ut)-\b(\vt))/2$ describe the
temporal
and spatial separation of the two points on the string world-tube with
coordinates $(u,v)$
and $(\ut,\vt)$ respectively.  The function $\psi$ is defined by
\begin{eqnarray}
  \psi(u,v,\ut,\vt) &=&
{1\over 8}[(\ap(u){\cdot}\ap(\ut)-1)(\bp(v){\cdot}\bp(\vt)-1)+\nonumber\\
&(&\ap(u){\cdot}\bp(\vt)-1)(\bp(v){\cdot}\ap(\ut)-1)-
    (\ap(u){\cdot}\bp(v)-1)(\bp(\vt){\cdot}\ap(\ut)-1)],
\end{eqnarray}
and $D$ is the linear differential operator
\begin{equation}
   D= U(u,v,\ut,\vt) \partial_u + V(u,v,\ut,\vt) \partial_v -
  \Ut (u,v,\ut,\vt) \partial_\ut - \Vt (u,v,\ut,\vt) \partial_\vt.
\end{equation}
The functions $U,V,\Ut$ and $\Vt$ are determined by requiring that $D$
satisfy the four equations
\begin{equation}
D\Delta t(u,v,\ut,\vt)=1,\qquad D\Delta {\bf x}(u,v,\ut,\vt)={\bf 0},
\label{Deq}
\end{equation}
which may be written as the matrix equation
\begin{equation}
\pmatrix{
1 & 1 & 1 & 1 \cr   &&&&\cr
\ap(u) & \bp(v) & \ap(\ut) & \bp(\vt) \cr}
\pmatrix{
U \cr V \cr \tilde U \cr \tilde V \cr}
=
\pmatrix{
2 \cr
{\bf 0} \cr}.
\label{matrix1}
\end{equation}
The generic solution to this set of equations is given in equation (3.12)
of \cite{AllenCasper}.

We restrict our attention to the limited class of loops described above
for the remainder of the paper.  With this restriction, (\ref{gamma1})
will lead directly to a simple formula for $\gamma$.  The function $\a(u)$
for this limited class of string loops may be written as
\begin{equation}
   {\bf a}(u) = \cases{u \k & {\rm for}\quad $u\in [0,\half)$, \cr &\cr
                       (1-u) \k & {\rm for}\quad $u \in [\half , 1)$, \cr}
\end{equation}
where $\k$ is the unit vector along the $z$-axis.  From this equation and
the fact that the $\b$-loop lies entirely in the plane perpendicular to
$\k$, it follows immediately that $\psi$ takes the form
\begin{equation}
  \psi(u,v,\ut,\vt) = \cases{ 0 &{\rm for}\quad$u\in [0,\half)$,
$\ut \in [0,\half)$
             \  or \ $u \in [\half,1)$, $\ut \in [\half,1)$ \cr &\cr
      {1 \over 4} \bigl( 1- \bp(v){\cdot}\bp(\vt) \bigr) & {\rm for}\quad
$u\in [0,\half)$,
     $\ut \in [\half,1)$\  or \ $u\in[\half,1)$, $\ut \in [0,\half)$.\cr}
\label{newpsi}
\end{equation}
In addition, for $u\in [0,\half)$ and $\ut \in [\half,1)$, equation
(\ref{matrix1})
for the coefficients in the differential operator $D$ becomes
\begin{equation}
\pmatrix{
1 & 1 & 1 & 1 \cr   &&&&\cr
\k & \bp(v) & -\k & \bp(\vt) \cr}
\pmatrix{
U \cr V \cr \tilde U \cr \tilde V \cr}
=
\pmatrix{2 \cr {\bf 0} \cr}
\end{equation}
with the obvious solution $U=\tilde U = 1$, $V=\tilde V =0$, i.e,
\begin{equation}
   D = \partial_u - \partial_\ut .
\label{newD}
\end{equation}
It is clear that the same operator also satisfies equation (\ref{Deq})
for $u\in [\half,1)$ and $\ut \in [0,\half)$.
(It is interesting that this solution works whether or not the
determinant of the matrix in (\ref{matrix1}) vanishes.)  Substituting
 (\ref{newpsi}) and (\ref{newD})
for $\psi$ and $D$ into (\ref{gamma1}), we find that
\begin{eqnarray}
\gamma&=& \idv  \int_{-\infty}^{\infty} d\vt\>
  \bigl( 1- \bp(v){\cdot}\bp(\vt) \bigr) \times \nonumber\\
    &&\Biggl\{ \int_0^\half du\>\int_\half^1 d\ut\>
       + \int_\half^1 du \> \int_0^\half d\ut \> \Biggr\}
    (\partial_u - \partial_\ut)  \Bigl[ \epsilon( \Delta t)
\delta\bigl( (\Delta t)^2-|\Delta {\bf x}|^2\bigr)\Bigr] .
\end{eqnarray}
For the specific form of ${\bf a}(u)$, the
second integral in curly braces immediately reduces to the first
under the change of variables $\ut=1- u'$, $u=1-{\tilde u}'$.
  In addition, we may perform the same
change of variables in the second term of the derivative operator.
Doing so leads directly to
\begin{equation}
\gamma=  4 \idv  \int_{-\infty}^{\infty} d\vt\>
  \bigl( 1- \bp(v){\cdot}\bp(\vt) \bigr) Q(v,\vt) ,
\label{gamma3}
\end{equation}
where
\begin{equation}
Q(v,\vt)=\int_0^\half du\>\int_\half^1 d\ut\>
      \partial_u \left[ \epsilon( \Delta t)
\delta( (\Delta t)^2-|\Delta {\bf x}|^2)\right].
\label{Q}
\end{equation}
The next step will be to carry out the integrations over $u$ and $\ut$
in $Q(v,\vt)$.

The partial derivative in (\ref{Q}) allows the integral over $u$ to be
performed immediately.  In doing so, the function in square brackets in
(\ref{Q}) must be evaluated at both the upper and lower limits of the
$u$ integration, yielding
\begin{equation}
Q(v,\vt)= \int_\half^1 d\ut\>\Big [\epsilon(\Delta t)\delta((\Delta t)^2
-|\Delta {\bf x}|^2)     \Big ]^{u=\half}_{u=0}\equiv T(v,\vt)-B(v,\vt),
\label{uint}
\end{equation}
where $T$ is the integral of the quantity in square brackets evaluated at
the top limit, and $B$ is the integral of the quantity in square brackets
evaluated at the bottom limit.  Consider first the top limit, $T(v,\vt)$.
Because $\ut \in [\half,1)$ we have
\begin{eqnarray}
    \Delta t &=& \half - \ut + v - \vt,\\
    \Delta {\bf x} &=& (\ut -\half)\k + {\bf b}(v) - {\bf b}(\vt) .
\end{eqnarray}
Introducing $u' = \ut - \half$, the integral over $\ut$ in $T(v,\vt)$ can be
 written as
\begin{equation}
    T(v,\vt)=\int_0^\half du' \> \epsilon (v-\vt- u')
            \delta\biggl( {1 \over 4} \Bigl[  - 2 u'(v-\vt) + (v-\vt)^2
             - \bigl({\bf b}(v) - {\bf b}(\vt) \bigr)^2 \Bigr] \biggr) .
\label{utint}
\end{equation}
Turning to the lower limit, $B(v,\vt)$, we have
\begin{eqnarray}
    \Delta t &=&  - \ut + v - \vt ,\\
    \Delta {\bf x} &=& (\ut - 1)\k + {\bf b}(v) - {\bf b}(\vt) .
\end{eqnarray}
We now use the freedom to rewrite (\ref{gamma3}) as an integral with $v$
running from $-\infty$ to $\infty$ and $\vt$ running from $0$ to $1$.
(This may readily be seen as follows.  Rewrite the integral over $\vt$ as
a sum over $n$ from $-\infty$ to $\infty$ of an integral from $0$ to $1$
of the integrand with $\vt$ replaced by $\vt + n$.  Then use the explicit
form of the integrand and the periodicity of $\bf b$ to combine the $n$
with the $v$ to transform the $v$ integral  into an integral from $-\infty$
to $\infty$.\cite{AllenCasper}) If we now make the simultaneous changes of
variable $\vt = v'$, $v = \vt'+1$ and $\ut = 1 - u'$ we recover exactly
(\ref{gamma3}) (with $v$ and $\vt$ replaced everywhere by $v'$ and $\vt'$)
except that the sign of the argument of the $\epsilon$-function has changed.
However, since the $\epsilon$-function is odd and we are dealing with the
bottom limit in (\ref{uint}), we recover exactly (\ref{utint}) for
$-B(v,\vt)$.  Hence, $Q(v,\vt)=2T(v,\vt)$.

Using (\ref{uint}), (\ref{utint}), and $T=-B$, $Q(v,\vt)$ may be rewritten
in the form
\begin{equation}
Q(v,\vt)={4 \over |v- \vt|} \int_0^\half du' \> \epsilon (v-\vt-u')
            \delta\Biggl( u' - {\Bigl[ (v-\vt)^2
             - \bigl({\bf b}(v) - {\bf b}(\vt) \bigr)^2 \Bigr]
                         \over 2(v-\vt)} \Biggr).
\end{equation}
This integral may be carried out immediately, yielding
\begin{eqnarray}
Q(v,\vt)=&&{4 \over |v- \vt|}
     \epsilon \Biggl({(v-\vt)^2 + \left({\bf b}(v) - {\bf b}(\vt) \right)^2
                         \over 2(v-\vt)} \Biggr) \times\nonumber\\
      &&\theta \Biggl({(v-\vt)^2 - \left({\bf b}(v) - {\bf b}(\vt) \right)^2
                         \over 2(v-\vt)} \Biggr)
 \theta \left(\half - {(v-\vt)^2 - \left({\bf b}(v) - {\bf b}(\vt) \right)^2
                         \over 2(v-\vt)} \right)  .
\label{utint2}
\end{eqnarray}
This answer may be considerably simplified.

First, because the numerator of the argument of the $\epsilon$-function
in $Q(v,\vt)$ is
manifestly positive, the function is just equal to $\epsilon(v-\vt)$.
This term simply has the effect of removing the absolute value signs from
the denominator of the prefactor.
Next, consider the two $\theta$-functions appearing in (\ref{utint2}):

(1) $|v-v'|$ is the arc-length along the $\b$-loop
from ${\bf b}(v)$ to ${\bf b}(\vt)$ (traversed in a particular sense
possibly more than once) which must
be greater than the straight line distance
$|{\bf b}(v) - {\bf b}(\vt)|$ between the two points.  Thus again the
numerator is a positive function and the first $\theta$-function is just
equal to $\theta(v-\vt)$.

(2) The argument of the second $\theta$-function can be written as
\begin{equation}
    {\bigl(1- (v-\vt)\bigr)(v-\vt) +
            \bigl({\bf b}(v) - {\bf b}(\vt) \bigr)^2
                         \over 2(v-\vt) } .
\end{equation}
We may neglect the denominator
which is always positive when the first $\theta$-function is non-vanishing.
 In addition, we need only consider the behavior of the numerator
for $v-\vt > 0$.
First we note that for $(v-\vt) \in (0,1)$ it is manifestly positive.  On
the other hand for $(v-\vt) > 1$, we note that $(v-\vt)$
is the arc-length along the $\b$-loop from ${\bf b}(v)$ to ${\bf b}(\vt)$
traversed in a particular sense (possibly more than once) while $(v-\vt)-1$
is the arc-length along the $\b$-loop from ${\bf b}(v)$ to ${\bf b}(\vt)$
traversed in the opposite sense (possibly more than once).  Thus $(v-\vt)
\geq |{\bf b}(v) - {\bf b}(\vt)|$ and
$(v-\vt)-1 \geq |{\bf b}(v) - {\bf b}(\vt)|$ and so
\begin{equation}
\bigl(1- (v-\vt)\bigr)(v-\vt) +
            \bigl({\bf b}(v) - {\bf b}(\vt) \bigr)^2 \leq 0 \qquad
              {\rm for}\quad (v-\vt) > 1 .
\end{equation}
Combining these results we see that (\ref{utint2}) simplifies to yield
\begin{equation}
Q(v,\vt)={4 \over (v-\vt)} \theta(v-\vt) \theta\bigl(1 - (v-\vt) \bigr).
\label{arg}
\end{equation}
Replacing $Q(v,\vt)$ in (\ref{gamma3}) with (\ref{arg}),
we obtain our principal result,
\begin{eqnarray}
    \gamma &=& 16 \int_0^1 dv\> \int_{v-1}^v d\vt \>
                   {1-\bp(v){\cdot}\bp(\vt) \over (v-\vt)}\nonumber\\
&&\nonumber\\
      &=& 16 \int_0^1 dv\> \int_0^1 dx \> {1-\bp(v){\cdot}\bp(v-x) \over x}.
\label{simplegamma}
\end{eqnarray}
In the next section we shall evaluate this remarkably simple expression
for a number of particular loops.

%%%%%%%%%%%%%%%%%%%%%%%%%%%%%%%%%%%%%%%%%%%%%%%%%%%%%%%%%%%%%%%%%%%%%%%%%%%%%%%

\section{RESULTS}
\label{section3}

\subsection{Circular $\b$-loop}

Our first application of formula (\ref{simplegamma}) for $\gamma$ is
the case where the $\b$-loop winds $M$ times around a perfect circle (see
Figure \ref{fig1}).  Here $M$ is any positive integer.  In this case,
the function $\b(v)$ may be written as
\begin{equation}
\b(v)={1\over 2\pi M}[\cos(2\pi M v)\i+\sin(2\pi M v)\j].
\end{equation}
The dot product appearing in (\ref{simplegamma}) gives
\begin{eqnarray}
\bp(v)\cdot\bp(v-x)&=&[\cos(2\pi M v)\cos(2\pi M(v-x))
+\sin(2\pi M v)\sin(2\pi M(v-x))]\nonumber\\
&=&\cos(2\pi M x).
\end{eqnarray}
Thus, $\gamma$ is given by
\begin{eqnarray}
\gamma&=&16 \int_0^1 dv\> \int_0^1 dx \> {1-\cos(2\pi Mx) \over x}
=16\int_0^1dx \> {1-\cos(2\pi Mx)\over x}\nonumber\\
&&\nonumber\\
&=&16\int_0^{2\pi M}dx{1-\cos x\over x}=16[C+\ln(2\pi M)-{\rm Ci}(2\pi M)]
\end{eqnarray}
where $C=0.577216\dots$ is Euler's constant and Ci$(x)$ is the cosine integral
function defined by equation (5.2.2) of reference \cite{AbramStegun}.
The numerical value of $\gamma$ for this string loop is $\gamma\approx
39.002454$ when $M=1$.
In section \ref{section4} we show that this is the minimum value of $\gamma$
for any loop in the limited class of loops for which the formula
(\ref{simplegamma}) is valid.

%%%%%%%%%%%%%%%%%%%%%%%%%%%%%%%%%%%%%%%%%%%%%%%%%%%%%%%%%%%%%%%%%%%%%%%%%%%%%%%

\subsection{Polygon $\b$-loop}

As the next application of formula (\ref{simplegamma}), we consider
 the case where the $\b$-loop takes the shape of a regular, $N$ sided
 polygon (see Figure \ref{fig2}).  Because the length of a given side
of the polygon is
 $1/N$, it is convenient to break the integrations in (\ref{simplegamma})
 into $N$ equally spaced intervals,
 \begin{equation}
 \gamma=16\sum\limits_{i,j=0}^{N-1}\int_{i/N}^{(i+1)/N}dv
\int_{j/N}^{(j+1)/N}  dx {1-\bp(v)\cdot\bp(v-x)\over x}.
 \end{equation}
For each $i$ and $j$, the $v$ and $x$ integrations are over specific
segments on the polygon.  The integrations may be simplified by
introducing a new pair of coordinates, $\vb$ and $\xb$ , which will each
be in the range $[0,1/N]$ :
\begin{equation}
\vb=v-{i\over N},\qquad \xb=x-{j\over N}.
\end{equation}
With this change of variables, $\gamma$ becomes
\begin{equation}
\gamma=16\sum\limits_{i,j=0}^{N-1}\int_0^{1/N}d\vb\int_0^{1/N}d\xb
{1-\bp(i/N+\vb)\cdot\bp(i/N+\vb-j/N-\xb)\over j/N+\xb}.
\end{equation}
The vector $\bp(i/N+\vb)$ is
a constant vector (tangent to the $i$th segment on the polygon) for each
value of $i$. When $\xb<\vb$, the vector $\bp(i/N+\vb-j/N-\xb)$ will be
the constant vector tangent to the $(i-j)$th segment on the polygon.
When $\xb>\vb$, it will be tangent to the $(i-j-1)$th segment.
Thus, the $\xb$ integration may be broken into two ranges where the
dot product $\bp(i/N+\vb)\cdot\bp(i/N+\vb-j/N-\xb)$ is constant in each
range.
\begin{eqnarray}
\gamma=16\sum\limits_{i,j=0}^{N-1}\int_0^{1/N}d\vb&&\bigg\{\int_0^{\vb}d\xb
{1-\bp(i/N+\vb)\cdot\bp(i/N+\vb-j/N-\xb)\over j/N+\xb}\nonumber\\
&&\nonumber\\
&+&\int_\vb^{1/N}d\xb{1-\bp(i/N+\vb)\cdot\bp(i/N+\vb-j/N-\xb)\over
j/N+\xb}\bigg\}
\label{pgam1}
\end{eqnarray}
For each value of $i$ and $j$, the constant dot products in
(\ref{pgam1}) may be evaluated.  Because $\bp(i/N+\vb)$ is tangent to
the $i$th segment on the polygon and $\bp(i/N+\vb-j/N-\xb)$ is tangent
to either the $(i-j)$th segment (for $\vb>\xb$) or the ($i-j-1$)th
segment (for $\xb>\vb$), it follows that their dot product is given by
\begin{equation}
\bp(i/N+\vb)\cdot\bp(i/N+\vb-j/N-\xb)=\cases{&$\cos({2\pi\over N}j)
\quad{\rm for}\>\vb>\xb$,\cr &\cr
&$\cos({2\pi\over N}(j-1))\quad{\rm for}\>\vb<\xb$,\cr}
\label{dots1}
\end{equation}
where $2\pi/N$ is the angle through which one must rotate a segment of
the polygon to bring it parallel to the next segment.
The r.h.s. of (\ref{dots1}) is independent of $i$.  Thus, the sum over $i$
simply gives a factor of $N$.  The integrations over $\xb$ may be done giving
\begin{eqnarray}
\gamma=16N\sum\limits_{j=0}^{N-1}\int_0^{1/N}d\vb\bigg\{&&
\Big(1-\cos({2\pi\over N}j)\Big)\Big(\ln(j/N+\vb)-\ln(j/N)\Big)\nonumber\\
&&+\Big(1-\cos({2\pi\over N}(j+1))\Big)\Big(\ln(j/N+1/N)
-\ln(j/N+\vb)\Big)\bigg\}.
\end{eqnarray}
The $\vb$ integration may now be carried out.  After combining terms,
this yields
\begin{eqnarray}
\gamma=16\sum\limits_{j=1}^{N-1}\bigg[\Big(&&1+j\cos({2\pi\over N}(j+1))
-(j+1)\cos({2\pi\over N}j)\Big)\ln\Big({j+1\over j}\Big)\nonumber\\
&&+\cos({2\pi\over N}j)-\cos({2\pi\over N}(j+1))\bigg]
+16\big(1-\cos({2\pi\over N})\big).
\label{pgam2}
\end{eqnarray}
Finally, we note that all the terms not multiplied by
$\ln\big({j+1\over j}\big)$ in (\ref{pgam2}) cancel exactly in the sum
over $j$.  Thus, we arrive at the final form
\begin{equation}
\gamma=16\sum\limits_{j=1}^{N-1}\Big(1+j\cos({2\pi\over N}(j+1))
-(j+1)\cos({2\pi\over N}j)\Big)\ln\Big({j+1\over j}\Big),
\label{polygamma}
\end{equation}
or equivalently
\begin{equation}
\gamma=32\bigg(1-\cos({2\pi\over N})\bigg)\bigg({1\over 2}N\ln N
+\sum_{j=2}^{N-1} j\ln (j)\> \cos({2\pi\over N} j)\bigg).
\label{polygamma2}
\end{equation}
For the first few values of $N$, equations (\ref{polygamma}) and
(\ref{polygamma2}) give
\begin{equation}
\gamma=\cases{&$64\ln 2\quad {\rm for}\quad N=2$,\cr &\cr
&$72\ln 3-48\ln 2\quad {\rm for}\quad N=3$,\cr &\cr
&$64\ln 2\quad {\rm for}\quad N=4$.\cr}
\end{equation}
Note that the $N=2$ case is identical to the Garfinkle
and Vachaspati case (equation (3.9) of reference \cite{GV}) with
$\alpha=\pi/2$, and has the same value of $\gamma$ as they obtained.
Numerical values for $\gamma$ are shown as a function of $N$ in
Figure \ref{fig3}.

It is interesting to note that (\ref{polygamma}) may be trivially modified
to obtain a formula for $\gamma$ for a set of (self-intersecting) loops
related to the
polygon case.  Consider again a $\b$-loop which is composed of $N$ straight,
equal length segments.  However, in this case, instead of placing each
segment at an angle of $2\pi/N$ relative to the previous segment (and
thereby getting an $N$ sided polygon), each successive segment is placed
at an angle of $2M\pi/N$ relative to the previous segment (where $M$ is
a positive integer not equal to $N$).  This causes
the $\b$-loop to wind  $M$ times around the origin.  The
only change required
in (\ref{polygamma}) is to replace the factor $2\pi/N$ appearing in the
cosine terms by $2M\pi/N$,
\begin{equation}
\gamma=16\sum\limits_{j=1}^{N-1}\Big(1+j\cos({2\pi M\over N}(j+1))
-(j+1)\cos({2\pi M\over N}j)\Big)\ln\Big({j+1\over j}\Big),
\label{mpgam}
\end{equation}
or equivalently
\begin{equation}
\gamma=32\bigg(1-\cos({2\pi M\over N})\bigg)\bigg({1\over 2}N\ln N
+\sum_{j=2}^{N-1} j\ln (j)\> \cos({2\pi M\over N} j)\bigg).
\end{equation}
If $M=1$, we have the
polygon case considered above.  If for instance, $M=2$ and $N=5$, the
$\b$-loop would have the shape of a pentagram (see Figure \ref{fig4}).

%%%%%%%%%%%%%%%%%%%%%%%%%%%%%%%%%%%%%%%%%%%%%%%%%%%%%%%%%%%%%%%%%%%%%%%%%%%%%%%

\subsection{Isosceles $\b$-loop}

Our final application of (\ref{simplegamma}) is the case
where the $\b$-loop takes the shape of an isosceles triangle.  This is
a one-parameter family of loops where the parameter is the half-angle
$\theta$ between the two equal length sides of the triangle (see Figure
\ref{fig5}).  The length of the two equal sides of the triangle is
denoted by $l_1$.
The length of the base of the triangle is denoted $l_2$.  Using the
constraint $2l_1+l_2=1$, one can easily write $l_1$ and $l_2$ in terms
of the semi-angle $\theta$ of the triangle,
\begin{equation}
l_1={1\over 2(\sin\theta+1)},\qquad l_2=2l_1\sin\theta.
\label{l1l2}
\end{equation}
The parameter along the $\b$-loop is taken to run from $0$ to $l_1$ along
the first side of the triangle, from $l_1$ to $2l_1$ along the second side,
and from $2l_1$ to $1$ along the third.  If we label the unit tangent
vectors to the first, second and third sides of the triangle by
$\bp_1, \bp_2$ and $\bp_3$ respectively, the dot products between the
various sides are given in terms of $\theta$ by
\begin{eqnarray}
\bp_1\cdot\bp_2&=&-\cos(2\theta),\nonumber\\
\bp_1\cdot\bp_3&=&\bp_2\cdot\bp_3=-\sin\theta.
\label{dots}
\end{eqnarray}
The dot product of any unit tangent vector with itself is simply $1$.

We are now in a position to use equation (\ref{simplegamma}) to find
$\gamma$ for this family of string loops.  The integrations over $v$
and $\vt$ may be broken into ranges corresponding to the three sides of
the triangle, yielding
\begin{eqnarray}
\gamma=16\Bigg\{&&\!\!\!\int_0^{l_1}dv\Bigg[
\int_{v-1}^{-l_2-l_1}d\vt{1-\bp_1\cdot\bp_1\over v-\vt}
+\int_{-l_2-l_1}^{-l_2}d\vt{1-\bp_1\cdot\bp_2\over v-\vt}\nonumber\\
&&\qquad\qquad\qquad\qquad
+\int_{-l_2}^0d\vt{1-\bp_1\cdot\bp_3\over v-\vt}
+\int_0^vd\vt{1-\bp_1\cdot\bp_1\over v-\vt}\Bigg]\nonumber\\
&+&\int_{l_1}^{2l_1}dv\Bigg[
\int_{v-1}^{-l_2}d\vt{1-\bp_2\cdot\bp_2\over v-\vt}
+\int_{-l_2}^0d\vt{1-\bp_2\cdot\bp_3\over v-\vt}\nonumber\\
&&\qquad\qquad\qquad\qquad
+\int_0^{l_1}d\vt{1-\bp_2\cdot\bp_1\over v-\vt}
+\int_{l_1}^vd\vt{1-\bp_2\cdot\bp_2\over v-\vt}\Bigg]\nonumber\\
&+&\int_{2l_1}^1dv\Bigg[
\int_{v-1}^0d\vt{1-\bp_3\cdot\bp_3\over v-\vt}
+\int_0^{l_1}d\vt{1-\bp_3\cdot\bp_1\over v-\vt}\nonumber\\
&&\qquad\qquad\qquad\qquad
+\int_{l_1}^{2l_1}d\vt{1-\bp_3\cdot\bp_2\over v-\vt}
+\int_{2l_1}^vd\vt{1-\bp_3\cdot\bp_3\over v-\vt}\Bigg]\Bigg\}.
\label{longgam}
\end{eqnarray}
The numerator of each integral in (\ref{longgam}) has the form
 $1-\bp_i\cdot\bp_j$.  Each of these numerators is now constant.  If
$i=j$, the numerator (and hence the integral) vanishes.  When $i\ne j$,
equation (\ref{dots}) may be used to write the dot products in terms
of $\theta$.  The integrals in (\ref{longgam}) are now easy to evaluate
and yield
\begin{eqnarray}
\gamma=-16\big\{&(&1+\cos(2\theta))(2(l_1+l_2)\ln(l_1+l_2)-l_2\ln(l_2)
+2l_1\ln(l_1)-2l_1\ln(2l_1))\nonumber\\
&+&(1+\sin\theta)(2l_1\ln(2l_1)+l_2\ln(l_2)
+l_2\ln(l_2)+2l_1\ln(2l_1))\big\}.
\end{eqnarray}
Using (\ref{l1l2}) to write $l_1$ and $l_2$ in terms of $\theta$ and
simplifying gives the final form
\begin{eqnarray}
\gamma=32\Big[&2&\ln 2\cos^2\theta-\sin^2(\theta)\ln(\sin\theta)
+(2-\sin\theta)(1+\sin\theta)\ln(1+\sin\theta)\nonumber\\
&&\nonumber\\
&-&(1-\sin\theta)(1+2\sin\theta)\ln(1+2\sin\theta)\Big].
\label{isosgam}
\end{eqnarray}
This formula for $\gamma$ is plotted in Figure \ref{fig6} for a range
of angles $\theta$.  It should be noted that (\ref{isosgam}) correctly
reduces to the $M=1,\>N=2$ result (\ref{mpgam}) of the previous section when
$\theta=0^{\circ}$ or $90^{\circ}$, and to the $M=1,\>N=3$ result when
$\theta=30^{\circ}$.

%%%%%%%%%%%%%%%%%%%%%%%%%%%%%%%%%%%%%%%%%%%%%%%%%%%%%%%%%%%%%%%%%%%%%%%%%%%%%%%

\section{MINIMUM $\gamma$ LOOP AND STATIONARY POINTS OF $\gamma$}
\label{section4}
In this section, we obtain the minimum value and stationary points of
$\gamma$ for {\it all} loops in our limited class.  Since ${\bf b}(v)$
is a periodic function on the interval $v \in [0,1)$, its derivative
w.r.t. $v$ is also a periodic function, and may be expressed as the
Fourier series
\begin{equation}
{\bf b}'(v) = \sum_{n=-\infty}^\infty e^{2 \pi i n u} \biggl(a_n \i
+ b_n \j \biggr).
\end{equation}
Since the function ${\bf b}'(v)$ is real, one has $a_n = a_{-n}^* $ and
$b_n = b_{-n}^* $ with ${}^*$ denoting complex conjugation.  The
periodicity of ${\bf b}$ also implies that $a_0$ and $b_0$ vanish,
since
\begin{equation}
\int_0^1 {\bf b}'(v) dv = {\bf b}(1)-{\bf b}(0) = a_0 \i +
b_0 \j={\bf 0}.
\end{equation}
For the remainder of this section, we use the notation $\sum'_n$ to
indicate the sum from minus infinity to infinity with the $n=0$ term
excluded.

The Fourier coefficients $a_n $ and $b_n$ are not arbitrary; they
satisfy an infinite number of constraints which follow from the gauge
condition that the length of the vector $| {\bf b}'(v) |  = 1$.  One
of these constraints may be
expressed in a useful ``integrated" form:
\begin{equation}
1 = \int_0^1 | {\bf b}'(v) |^2 dv = \int_0^1 \sumprime_n \sumprime_m
e^{2 \pi i ( n+m) v } (a_n a_m + b_n b_m) dv \\
= \sumprime_n  (a_n a^*_n + b_n b^*_n)
\label{constraint}
\end{equation}
where in the final line we have used the orthogonality of the
exponential functions on the unit interval and the reality conditions
$a_n = a_{-n}^* $ and $b_n = b_{-n}^* $.  One may define a non-negative
real quantity $c_n^2 = 2 (a_n a^*_n + b_n b^*_n)$.  Note that for any
cosmic string loop one has $c^2_n = c^2_{-n}$, and that the normalization
condition
\begin{equation}
\sum_{n=1}^\infty c_n^2 = 1
\end{equation}
is implied by the integrated constraint (\ref{constraint}).

Now consider the value of $\gamma$ given by equation (\ref{simplegamma}),
\begin{eqnarray}
\gamma & =  &16 \int_0^1 dv \int_0^1 {dx \over x} \>
 \bigl[ { 1-\bp(v)  \cdot \bp(v-x) \bigr]} \nonumber\\
& =  &16 \int_0^1 {dx \over x} \> \int_0^1 dv
\> \biggl[ { 1- \sumprime_n \sumprime_m
e^{2 \pi i ((n + m)v - mx)} (a_n a_m + b_n b_m) } \biggr] \nonumber\\
& = &16 \int_0^1 {dx \over x}  \> \biggl[ { 1- \sumprime_n
e^{2 \pi i n x} (a_n a^*_n + b_n b^*_n) } \biggr].
\label{firstgammaform}
\end{eqnarray}
For any cosmic string loop, the function $\b$ satisfies the
integrated constraint (\ref{constraint}). Substituting this for the
{\it first} term in the
previous equation (\ref{firstgammaform}), one obtains
\begin{equation}
\gamma = 16 \sumprime_n \int_0^1 dx {  \bigl( 1-
e^{2 \pi i n x} \bigr)  \over x} \> (a_n a^*_n + b_n b^*_n)
= \sum_{n=1}^\infty \lambda_n c^2_n.
\label{quadraticform}
\end{equation}
This is a quadratic form for $\gamma$; its ``eigenvalues" are
\begin{eqnarray}
\lambda_n & \equiv &16\Re \biggl[\int_0^1 {(1-e^{2\pi i n x})\over x}\>
dx \biggr] = 16 \int_0^{2 \pi n} {( 1-\cos x ) \over x} \> dx \nonumber\\
& = & 16 \bigl[ C + \ln(2 \pi n) - {\rm Ci}(2 \pi n ) \bigr] \nonumber\\
& = & 16 \bigl[ C + \ln(2 \pi n) - \int_0^\infty {t e^{-2 \pi n t}
\over 1 + t^2} dt \> \bigr] \nonumber\\
&\approx & 16 \bigl[  C + \ln(2 \pi n) - {1 \over 4 \pi^2 n^2} +
O({1 \over n^4}) \> \bigr],
\end{eqnarray}
where $C = 0.577216...$ is Euler's constant and ${\rm Ci}(x)$ is the
cosine integral function defined by equation (5.2.2) of reference
\cite{AbramStegun}.
{}From the integral forms it is clear that the eigenvalues increase
without bound: $\lambda_n < \lambda_{n+1}$.  The first few eigenvalues
are approximately
\begin{equation}
(\lambda_1,\lambda_2,\lambda_3,\cdots) \approx
(39.0025,49.8297,56.2636,60.8473,64.4086,67.3208,...).
\end{equation}
{}From the quadratic form of $\gamma$ (\ref{quadraticform}) one may easily
find the absolute minimum and stationary points of $\gamma$.

Since any cosmic string loop in our limited class has
$\sum_{n=1}^\infty c_n^2 = 1$ and $\gamma = \sum_{n=1}^\infty c_n^2
\lambda_n$, it is immediately clear that the minimum value of $\gamma$
is attained when $c_1^2=1$ and all other $c^2_{n \ne 1}$ vanish.  (The
problem is identical to that of minimizing the energy of a quantum
mechanical system with orthonormal eigenstates $|n>$, where the general
normalized state is denoted $\sum_n c_n |n>$ and the energy of the
$n$'th state is $\lambda_n$.) The loop with $c_1^2=1$ is precisely the
$M=1$ case considered in Section 3A, where the ${\bf b}$-loop is a circle
traversed once as $v$ increases from zero to one.  The remaining
stationary points of $\gamma$ are immediately seen to be loops with
$c^2_k=1$ for some integer $k$ and all other coefficients vanishing.
This corresponds to the $M=k$ case considered in Section 3A where the
$\b$-loop is a circle traversed $k$ times as $v$ increases from zero to one.

%%%%%%%%%%%%%%%%%%%%%%%%%%%%%%%%%%%%%%%%%%%%%%%%%%%%%%%%%%%%%%%%%%%%%%%%%%%%%%%

\section{conclusion}
\label{conclusion}

In this paper we have found a simple formula for $\gamma$, the power
radiated in gravitational waves, for a limited class
of cosmic string loops.  We have applied this formula to a variety of
loop configurations within the limited class and have shown that the
minimum value for $\gamma$ in this class is attained only when the
$\b$-loop has the shape of a perfect circle (traversed once).

The results for $\gamma$ in the piecewise-linear cases (polygon and
isosceles $\b$-loops) have been checked against the general analytic
formula given in reference \cite{AllenCasper}, using a symbolic
manipulator ( Mathematica ) to evaluate the formula of reference
\cite{AllenCasper}.  We have also used these exact results to test a
floating-point implementation (written in C) of the exact formula of
reference \cite{AllenCasper}.  This floating-point implementation
was described
in some detail in \cite{AllenCasper} and is publicly available.  The
tests have shown that the floating-point implementation typically
yields results accurate to between 6 and 8 decimal places, the
difference being due to accumulated round-off error.

In reference \cite{AllenCasper} we presented an analytic formula
yielding $\gamma$ for piecewise-linear loops, and argued that any
smooth loop could be well-approximated by an $N$-segment piecewise-linear
loop with $N$ sufficiently large.  The exact formula in this paper allows
us to
determine the rate of convergence for large $N$, at least for a limited
class of loops.  The results for the circular $\b$-loop case considered
in section \ref{section3}A (with $M=1$) may be used to examine how well
smooth loops
are approximated by piecewise-linear loops.  The regular polygon
$\b$-loops ($M=1$) of section \ref{section3}B are increasingly circular
in shape as $N\rightarrow\infty$.  The difference between the $\gamma$
value for a cosmic string loop with a $\b$-loop in the shape of an $N$
sided polygon ($\gamma_N$) and the result from the case where the
$\b$-loop is a perfect circle ($\gamma_\infty$) is shown as a function
of $N$ in Figure \ref{fig7}.  The piecewise-linear result quickly
converges to the smooth result (see also Figure \ref{fig3}).  The
convergence goes like $1/N^2$ for $N > 10$.

\noindent
\vskip 0.3in
\centerline{ACKNOWLEDGMENTS}
\vskip 0.05in
This work was supported in part by NSF Grant No. PHY-91-05935 and a NATO
Collaborative
Research Grant.

%%%%%%%%%%%%%%%%%%%%%%%%%%%%%%%%%%%%%%%%%%%%%%%%%%%%%%%%%%%%%%%%%%%%%%%%%%%%%%%

%%%%%%%%%%%%%%%%%%%%%%%%%%%%%%%%%%%%%%%%%%%%%%%%%%%%%%%%%%%%%%%%%%%%%%%%%%%%%%%

\begin{figure}
\caption{A 2-segment $\a$-loop perpendicular to a circular $\b$-loop with
$M=1$.  The cosmic string loop defined by these $\a$- and $\b$-loops has
$\gamma\approx 39.002454$.  This is the minimum value of $\gamma$ attained
by any loop in the limited class which we consider.}
\label{fig1}
\end{figure}

\begin{figure}
\caption{A 2-segment $\a$-loop perpendicular to a $\b$-loop in the shape
of a regular polygon with $M=1$ and $N=5$. The cosmic string loop defined
by these $\a$- and $\b$-loops has $\gamma\approx 43.100206$. }
\label{fig2}
\end{figure}

\begin{figure}
\caption{Values of $\gamma$ for cosmic string loops with 2-segment
$\a$-loops perpendicular to $\b$-loops in the shape of regular $N$ sided
polygons with $M=1$.  These $\gamma$ values converge like $1/N^2$ to the
value of $\gamma$ for the circular $\b$-loop case.}
\label{fig3}
\end{figure}

\begin{figure}
\caption{A 2-segment $\a$-loop perpendicular to a $\b$-loop in the shape
of a regular polygon with $M=2$ and $N=5$. The cosmic string loop defined
by these $\a$- and $\b$-loops has $\gamma\approx 56.980278$.}
\label{fig4}
\end{figure}

\begin{figure}
\caption{A 2-segment $\a$-loop perpendicular to a $\b$-loop in the shape
of an isosceles triangle with semi-angle $\theta$.  The semi-angle $\theta$
lies in the range $\theta\in[0,\pi/2]$.}
\label{fig5}
\end{figure}

\begin{figure}
\caption{Values of $\gamma$ plotted as a function of the semi-angle
$\theta$ for the cosmic string loops described in Figure 5.  The cases
$\theta=0$ and $\theta=90^\circ$ yield the same result,
$\gamma=64\ln 2\approx 44.361419$, as Garfinkle and Vachaspati equation
(3.9) of reference \protect\cite{GV} with $\alpha=\pi/2$.}
\label{fig6}
\end{figure}

\begin{figure}
\caption{The convergence of $\Delta\equiv \gamma_N-\gamma_\infty$, where
$\gamma_N$ is the power radiated by a cosmic string loop with a
2-segment $\a$-loop perpendicular to an $N$ sided regular polygon
($M=1$) $\b$-loop, and $\gamma_\infty$ is the power radiated when
the $\b$-loop is a circle ($M=1$).  The value of $\Delta$ converges to zero
like $1/N^2$ for large $N$.}
\label{fig7}
\end{figure}

\end{document}